\documentclass[structabstract]{aa}  
\usepackage{graphicx}
\usepackage{txfonts}
\usepackage{natbib}
\bibpunct{(}{)}{;}{a}{}{,} % to follow the A&A style
\def\kmps{km\,s$^{-1}$}
\def\lxbol{$L_{\rm X,bol}$}

\def\flam{ergs\,cm$^{-2}$\,s$^{-1}$\,\AA$^{-1}$}
\def\lum{ergs\,s$^{-1}$}
\def\fint{ergs\,cm$^{-2}$\,s$^{-1}$}
\def\crat{s$^{-1}$}
\def\msun{M$_\odot$}

\begin{document}

\title{XMM-Newton observations of the low-luminosity cataclysmic variable 
V405 Pegasi\thanks{Based on observations obtained with XMM-Newton, an 
ESA science mission with instruments and contributions directly funded by 
ESA Member States and NASA}}

\author{A.D. Schwope\inst{1}
\and
V. Scipione\inst{1}
\and
I. Traulsen\inst{1}
\and
R. Schwarz\inst{1}
\and
T. Granzer\inst{1}
\and
A. M. Pires\inst{1}
\and
J.R. Thorstensen\inst{2}}

\institute{Leibniz-Institut f\"ur Astrophysik Potsdam (AIP),
              An der Sternwarte 16, 14482 Potsdam, Germany
              \and
              Department of Physics and Astronomy, 6127 Wilder Laboratory,
              Dartmouth College, Hanover, NH 03755-3528
}

\authorrunning{A.D. Schwope et al.}
\titlerunning{XMM-Newton observations of V405 Peg}
   
\date{Received 13 September 2013; accepted 28 November 2013}

\abstract{}{}{}{}{} 
% 5 {} token are mandatory
% context, aim, methods, results, conclusions 
\abstract
{V405 Peg is a low-luminosity cataclysmic variable (CV) that was identified as the
optical counterpart of the bright, high-latitude ROSAT all-sky survey source 
RBS1955. The system was suspected to belong to a largely undiscovered population 
of hibernating CVs. Despite intensive optical follow-up its subclass however remained 
undetermined.}
{We want to further classify V405 Peg and understand its role in the CV zoo via
its long-term behaviour, spectral properties, energy distribution and accretion luminosity.}
{We perform a spectral and timing analysis of \textit{XMM-Newton} X-ray and 
ultra-violet data. Archival WISE, HST, and Swift observations are used to determine 
the spectral energy distribution and characterize the long-term variability. } 
{The X-ray spectrum is characterized by emission from a multi-temperature plasma.
No evidence for a luminous soft X-ray component was found. Orbital phase-dependent 
X-ray photometric variability by $\sim$50\% occurred without significant spectral changes. 
No further periodicity was significant in our X-ray data. The average X-ray luminosity during 
the XMM-Newton observations was 
$L_{\rm X, bol} \simeq 5\times 10^{30}$\,\lum\ 
but, based on the Swift observations, the corresponding luminosity
varied between $5\times 10^{29}$\,\lum\ and $2\times 10^{31}$\,\lum\ on timescales of years. 
} 
{\rm The CV subclass of this object remains elusive. The spectral and timing properties 
show commonalities with both classes of magnetic and non-magnetic CVs. 
The accretion luminosity is far below than that expected for a standard accreting 
CV at the given orbital period.
Objects like V405 Peg might represent the tip of an iceberg and thus may be important contributors 
to the Galactic Ridge X-ray Emission. If so they will be uncovered by future X-ray surveys, e.g.~with eROSITA.
} 

\keywords{X-rays - stars: individual: V405 Peg - stars: Cataclysmic variables}

\maketitle
%
%________________________________________________________________

\section{Introduction}

Cataclysmic variables (CVs) are close interacting binaries, consisting of a white dwarf primary star
accreting matter from a low-mass late-type main-sequence secondary star via Roche lobe overflow. 
Depending on the strength of the magnetic field of the white dwarf an accretion disk may form or not.
In quiescent non-magnetic CVs, X-rays are thought to originate from a boundary layer between the 
inner disk and the surface of the white dwarf. In magnetic CVs X-rays originate from an accretion 
column at or near the magnetic poles. The columns may be fed via 
magnetospheric accretion from the inner boundary of a truncated disk, as in the subclass 
of Intermediate Polars (IPs), or from a funneled accretion stream in AM Herculis stars (also 
termed Polars). The X-ray spectra 
of magnetic and non-magnetic CVs are typically described by 
multi-temperature thermal plasma emission, being 
either collisionally or photo-ionized \citep[for reviews 
of the X-ray emission of magnetic and non-magnetic CVs see e.g.][]{baskill+05,mukai+03,kuulkers+06}.

The X-ray signal may be modulated on the orbital and the spin phase.
The X-ray brightness may show phase-dependent changes due to geometric obscuration 
by the stellar bodies (eclipses and self-eclipses) and by absorbing matter close to the 
accretion regions, to the disk or in the magnetosphere. 
X-ray reflection may occur in magnetic and non-magnetic CVs,
most obviously via the detection of the FeI $K\alpha$ line at 6.4\,keV. 

A distinct spectral feature of many magnetic CVs is a luminous soft X-ray component 
that can be described by blackbody emission in the range $kT \simeq 20-50$\,eV. 
The prevalence of this component led to the numerous discoveries of magnetic
CVs in the RASS (ROSAT all-sky survey) \citep[see e.g.][]{beubur95, schwope+02}.

V405 Pegasi (also known as RBS 1955) was identified as a bright, nearby CV 
in an optical identification program of bright, high-latitude RASS sources, 
the ROSAT Bright Survey \citep[RBS, ][]{schwope+00} and initially classified 
as a non-magnetic CV \citep{schwope+02}. Extensive optical spectroscopic 
and photometric follow-up observations obtained between 2002 and 2006 
were presented by \citet{thor+09}. This study revealed a parallax of $149^{+26}_{-20}$\,pc
and the occurrence of high and low states. The M3.5 ZAMS secondary star can clearly 
be recognized in the optical spectrum in all accretion states and dominates the photometric 
variability through ellipsoidal modulations in the low state. 
From M-star absorption-line spectroscopy an orbital period of 4.2635\,hr and a 
radial velocity semi-amplitude of $92\pm3$\,\kmps\ were obtained, 
implying a rather low orbital inclination if masses were typical. 
From the same data the time of inferior conjunction of the secondary 
star was measured, $T_0 = \mbox{HJD} 2452623.6767(1)$, which determines 
phase zero throughout this paper.

An unambiguous classification of the object was not possible.
Following \citet{thor+09} the occurrence of low states can be reconciled with 
a magnetic system. The object was found to be very much under-luminous at the given orbital period. 
This gave reason to speculate that V405 Peg might belong to a largely undiscovered 
population of hibernating CVs. Such a population of post-nova CVs was predicted to exist
to resolve the discrepancy between the observed and the predicted nova rates  \citep{shara+86}.

In order to better constrain its nature, hence CV subclass,
we performed X-ray and ultra-violet observations with XMM-Newton. 
Those observations were undertaken with the primary aim to establish 
its X-ray spectral parameters, to search for a soft component, 
and to search for the spin period of the accreting white dwarf. 

Our analysis is supported by several archival X-ray observations performed 
with the Swift-XRT and UVOT, and WISE infrared and HST ultra-violet observations. 

%__________________________________________________________________

\section{Observations and data reduction}

A summary of the observations reported in this paper is given in Table~\ref{t:log}.

\begin{table}[t]
\caption{Summary of new X-ray, ultra-violet and optical observations of V405 Peg\label{t:log}}
\begin{tabular}{llrl}
\hline
Observatory & Start date & Exposure & Filter \\
\hline
XMM-Newton \\ 
0604060101\\
EPIC & 2009-12-29 &  54 ks & thin\\
OM & 2009-12-29 & 19.7 ks & UVM2 \\ 
OM & 2009-12-30 & 27.3 ks & UVW2 \\ 
\\
Swift \\
00037668001 & 2008-06-04 &1288 s& W2\\
00037668002 & 2008-07-25 &2156 s& U\\
00037668003 & 2008-07-29 &2500 s& U\\
00045762001 & 2012-10-18 &867 s& V,B,U,W1,M2,W2\\
00045762002 & 2012-10-21 & 513 s& W2\\
00045762003 & 2012-10-23 & 1186 s& V,B,U,W1,M2,W2\\
00045762004 & 2012-10-30 & 2091 s& V,B,U,W1,M2,W2\\
\\
STELLA \\ 
& 2012-07-12 & $97 \times 120$\,s & g \\
& 2012-07-13 & $127 \times 120$\,s & g \\
& 2012-07-14 & $123 \times 120$\,s & g \\
& 2012-07-15 & $131 \times 120$\,s & g \\
\hline
\end{tabular}
\end{table}

\subsection{XMM-Newton X-ray and ultraviolet observations}  
V405 Peg was observed with XMM-Newton for about $\sim$53\,ks on 2009 December 29 
(observation ID 0604060101), thus covering 3.45 cycles of the 4.26\,h binary. 
The EPIC-pn and -MOS cameras were operated in full frame 
mode with the thin filter inserted. The RGS spectrographs were used with their spectroscopy modes, respectively.
Simultaneous ultraviolet photometry was obtained with the Optical Monitor (OM) in timing mode. 
The \textit{UVM2} and \textit{UVW2} filters were used sequentially, the UVM2 for $\sim$22.7\,ks (1.5 binary cycles) and the UVW2 for $\sim$29.0\,ks (1.9 binary cycles. The filters cover the wavelength intervals 1970--2675\,\AA\ and 1805--2454\,\AA, respectively.

The raw-data were reduced using the \textit{XMM-Newton} Science 
Analysis System (\textsc{sas}) version 11.0.0.
The EPIC-pn and MOS data were processed with the standard tasks \textit{epproc} and \textit{emproc}, respectively,
to generate calibrated event lists.
The observations were affected during the first 12.5\,ks of the XMM-Newton observations
by non-optimum space weather, with background count rates in excess of 0.4\,s (EPIC-pn, 
$0.2 - 12$\,keV).
For the spectral and timing analysis those time intervals were removed.

For both the MOS and pn detectors, spectra were extracted from a circular source region of 
45\,arcsec\ radius and light curves from a circular source region of 25\,arcsec\ radius, 
and the background from either a circular or a box-shaped source-free region on the same CCD.
Different extraction regions were used for the spectral and the timing analyses due to 
the different impact on the unstable background signal in the two cases. The count rate accuracy 
of individual time bins is more affected by an unstable background than the average spectrum.

The response files for the spectral analysis were generated using the \textit{rmfgen} and 
\textit{arfgen} tasks and all spectra were rebinned to a minimum of 20 counts per bin in 
order to use the $\chi^2$ statistics for goodness of fit and parameter estimation.
The mean net count rates were $0.627 \pm 0.004$\,s$^{-1}$ for EPIC-pn, 
and $0.160 \pm 0.002$\,s$^{-1}$ for EPIC-MOS (mean of both cameras).

All timing data shown in this paper were corrected to the barycenter of 
the Solar system with the \textit{barycen} task.  Background-subtracted light curves 
in different energy bands were generated using the task \textit{epiclccorr} with time bins of 100\,s 
and 200\,s. 
%----------------------------------------------------------- 
\begin{figure}[t]
\resizebox{\hsize}{!}{\includegraphics[clip=]{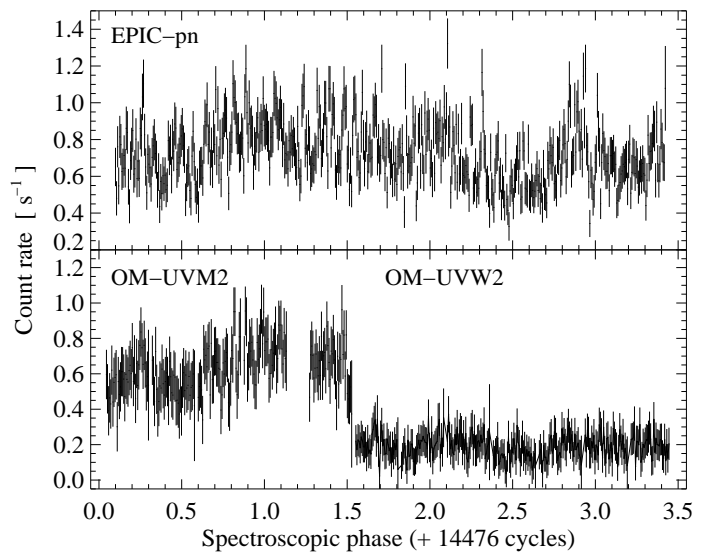}}
\caption{Background-subtracted X-ray and UV light curves of V405 Peg obtained 
with XMM-Newton on Dec.~29, 2009. The data were binned using a 
bin size of 100 s. They are shown in original time sequence with time 
transformed to binary phase using the ephemeris of \citet{thor+09}.
Phase zero refers to the inferior conjunction of the secondary star. }
\label{f:lcori}
\end{figure}

The RGS data were reprocessed with the standard task \textit{rgsproc}, by which 
first order source and background spectra were produced with mean net count rates of 
$0.013 \pm 0.001$\,s$^{-1}$ for RGS1 and $0.016\pm 0.001$\,s$^{-1}$ for RGS2.
Since the count rates were so low the RGS-data were not considered for the spectral analysis.

\begin{figure}[t]
\resizebox{\hsize}{!}{\includegraphics[clip=]{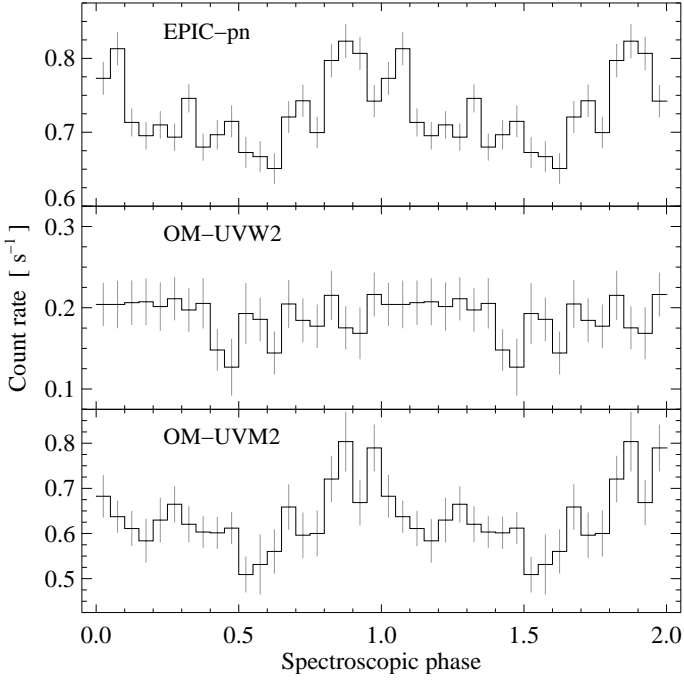}}
\caption{Phase-averaged XMM-Newton X-ray and UV light curves of V405 Peg 
with a phase resolution of 0.05 phase units. All data are shown twice for easier
recognition of repeated features.}         
\label{f:lcfol}
\end{figure}

The OM data were reduced and background-subtracted using the \textit{omfchain} 
routine with 100 s time binning. Barycentric corrections were also applied as for the EPIC data. 
The mean count rates were $0.63\pm 0.05$\,\crat\ in the \textit{UVM2} band and 
$0.19\pm0.03$\,\crat, in the \textit{UVW2} band. 
The light curves obtained with the optical monitor are shown in
original time sequence and phase-folded in Figs.~\ref{f:lcori} and \ref{f:lcfol}. 

%______________________________________________________________
\begin{figure}
\resizebox{\hsize}{!}{\includegraphics[clip=]{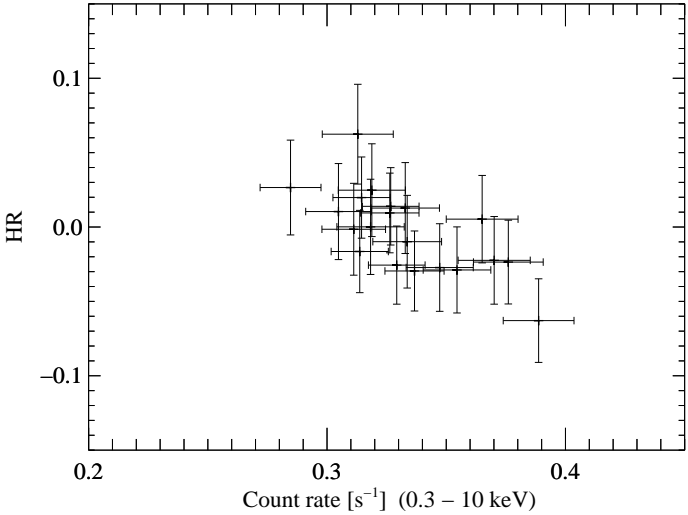}}
\caption{Hardness ratio vs.~soft count-rate of the phase-averaged EPIC-pn data in 20 phase bins.
For definitions of soft and hard bands see text.
}
\label{f:crhr}
\end{figure}

\begin{table}
\caption{Swift XRT parameters of V405 Peg. The count-rate was converted 
to bolometric flux assuming the same multi-temperature plasma emission 
model for all observations (see Sect.~\ref{s:swspec} for details)\label{t:xrat}}
\begin{tabular}{lcrc}
\hline
Observation & rate & $n_{\rm photons}$ & $F_{\rm bol}$\\
& s$^{-1}$ &  & \fint\\
\hline
00037668001 & $0.004 \pm 0.002$ & 5 & $2\times10^{-13} $\\
00037668002 & $0.029 \pm 0.004$ & 64 & $1.5\times10^{-12} $ \\
00037668003 & $0.043 \pm 0.004$ & 108  & $2.2\times10^{-12} $\\
00045762001 & $0.16  \pm 0.01$  & 140  & $8.0\times10^{-12} $\\
00045762002 & $0.09  \pm 0.01$  & 50  & $4.5\times10^{-12} $\\
00045762003 & $0.15  \pm 0.01$  & 181  & $7.5\times10^{-12} $\\
00045762004 & $0.149 \pm 0.008$ & 313  & $7.5\times10^{-12} $\\
\hline
\end{tabular}
\end{table}

\subsection{Swift XRT and UVOT observations}
The Swift observatory observed the field of V405 Peg in three occasions in July 2008 
and four occasions in October 2012 (see Table~\ref{t:log}). All observations were short, the
maximum observation length was about 2500\,s (16\% of the orbital cycle). 

The XRT-data were reduced with the online 
analysis tools provided by the ASI Science Data Center\footnote{http://www.asdc.asi.it}. 
Calibrated count rates, corrected for deadtime and coincidence losses, were read 
from the standard pipeline (level 3) products made available via the same web-site. 
Mean count-rates per observation obtained with the X-ray telescope XRT 
are listed in Table~\ref{t:xrat}. The results of UVOT photometry, kindly made available 
by Dr.~M.J.~Page (MSSL), are not listed in tabular form but shown graphically in Fig.~\ref{f:uvirsed}.  

The object was found to be strongly variable between the various occasions. At X-ray wavelengths the 
source varied by a factor 40 on timescales of years (between 2008 and 2012), by a factor of  
$\sim$10 on a timescale of weeks (between the observations obtained in 2008)
and by a factor of 2 within three days in October 2012.

The UV variability was found to be similarly pronounced. The largest amplitude 
(factor 45) is observed in the blue-most filter (UVW2) between 2008 and 2012.

\begin{figure}[t]
\resizebox{\hsize}{!}{\includegraphics[clip=]{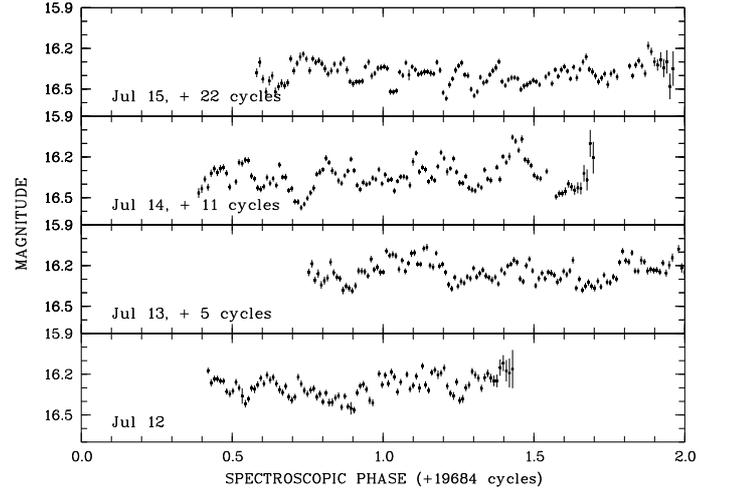}}
\caption{STELLA/WiFSIP differential photometry obtained in July 2012 
using an SDSS g-filter.}
\label{f:olcs}
\end{figure}

\subsection{STELLA/WiFSIP optical photometry}
V405 Peg was observed during four nights with the robotically controlled telescope STELLA-I 
located at an altitude of 2400\,m on the island of Tenerife \citep{strassmeier+04}. 
The telescope is permanently equipped 
with the Wide-Field STELLA Imaging Photometer \citep[WiFSIP, ][]{weber+12}.  Our data were
obtained during commissioning of the instrument through a $g$-filter. Individual exposure
times were 120\,s, the achieved time resolution with the available instrumental 
setup at that time was 160\,s. 
During each of the nights full phase coverage could be achieved. All observations were 
performed under clear sky with little seeing and transparency variations. 
Differential photometric magnitudes were computed with respect to the nearby object 
SDSS\,J230944.80$+$213537.2, the resulting light curves are shown in Fig.~\ref{f:olcs}.

\subsection{Further archival data on V405 Peg\label{s:aobs}}
The mean low-state spectrum obtained by \citet{thor+09} has 
synthetic magnitudes of $gri = 17.9, 16.5, 15.4$, respectively.

An SDSS spectrum was obtained on August 16, 2008. It shows a rich emission-line spectrum 
very similar to that shown by \citet{thor+09}, thus indicating a high accretion state. 
By folding the observed spectrum through the SDSS ugriz filter curves, approximate SDSS 
magnitudes were determined: $ugriz = 16.71, 16.47, 15.80, 14.84, 14.09$.
SDSS photometry, $ugriz = 15.78, 16.10, 15.35, 14.65, 14.09$, was indicative of another
even higher state of accretion (see Fig.~\ref{f:uvirsed} for a 
graphical representation of both, SDSS spectroscopy and photometry, and the mean low-state 
spectrum). 

Archival data are available in the infrared (2MASS, WISE) and the ultraviolet spectral ranges.
The 2MASS JHK magnitudes are 12.67, 12.01, and 11.81 ($\pm 0.02$\,mag for all three filters), 
and the WISE W1, W2, W3 magnitudes at central wavelengths  3.4, 4.2, and 12.0 $\mu$m are
$11.771\pm0.023, 11.608\pm0.022, 11.302\pm0.146$, respectively. 
The data in W4 are unreliable.

An HST spectrum was obtained on Dec.~7, 2012, with the Cosmic Origins Spectrograph (proposal ID 12870, 
PI B.~G\"ansicke) and grating G140L covering the wavelength range 1165 -- 2148\,\AA. 
Time-tagged data were obtained for about two hours of total exposure time. We make use of the 
time-averaged preview spectrum when discussing the spectral energy distribution in Sect.~\ref{s:sed}.
%_______________________________________________________________________________________________________________

\section{Analysis and results}
\subsection{X-ray spectral and timing analysis}
\subsubsection{\label{sss:xuvlc} XMM-Newton X-ray and ultraviolet light curves}

\begin{figure}[t]
\resizebox{\hsize}{!}{\includegraphics[bbllx=2pt,bblly=1pt,bburx=333pt,bbury=251pt,clip=]{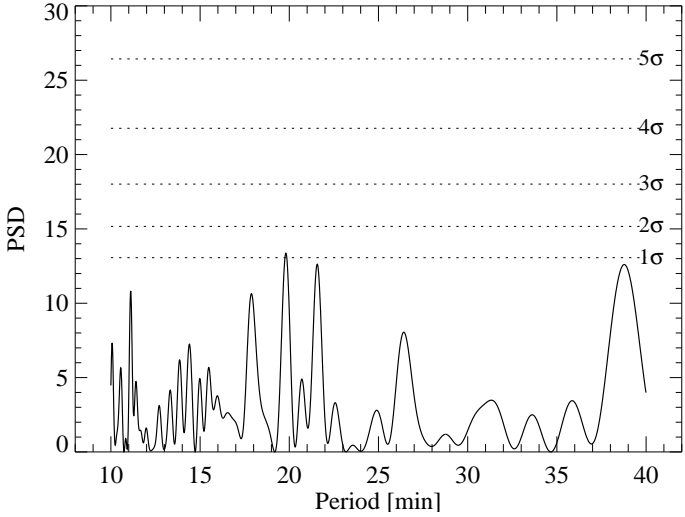}}
\caption{Lomb-Scargle periodogram of de-trended EPIC-pn data (boxcar mean subtracted).}
\label{f:scargle}
\end{figure}

X-ray light curves covering the whole EPIC energy range (0.2 - 12 keV) are 
shown in original time sequence and phase-averaged in Figs.~\ref{f:lcori} and \ref{f:lcfol}. 
The light curve in original time sequence does not show a clear orbital modulation but 
is affected by strong flickering. At a time resolution of 100\,s the EPIC-pn X-ray count rate
varied between 0.4 and 1.2\,\crat.  

When folded and averaged over the orbital phase of 4.26 hours the X-ray light curve shows
some weak orbital phase dependent modulation with bright and faint phases lasting from about
$\phi = 0.6$ to $1.15$, and from $\phi = 0.15$ to $0.60$, respectively. The mean count rates during 
those orbital phase intervals differ by about 15\%. The orbital minimum occurs at phase 0.6.

We searched for phase-dependent changes of the X-ray color by analysing the hardness 
ratio HR\footnote{The hardness ratio HR is defined as $HR = \frac{H-S}{H+S}$ with 
$H, S$ being the source counts in suitably chosen hard and soft spectral bands, respectively.} 
of the EPIC-pn data.
The whole energy band was divided in two sub-bands, $0.3-1.0$\,keV and $1-10$\,keV,
with approximately the same number of photons. The corresponding binned light curves 
with time bin size of 200\,s were found to resemble each other.
The hardness ratio does show a slight dependency on the orbital phase, the spectrum
being softest at highest count-rate. This behavior is best visible in the CR-HR 
plane shown in Fig.~\ref{f:crhr}. 

V405 Peg was observed with the OM-UVM2 for nearly 1.5 orbital cycles, and with the 
UVW2 for about 2 orbital cycles. While some variability is apparent in the UVM2 data, the UVW2 light
curve appears almost flat. The phase-folded UVM2 light curve resembles that of EPIC-pn, but 
part of this kind of variability with bright and faint phases 
might be mimicked by overall trends in the light curve. The overall
coverage of the system is too short to make a robust statements about a possible phase-dependent 
UV variability.

\begin{figure}[t]
\resizebox{\hsize}{!}{\includegraphics[clip=]{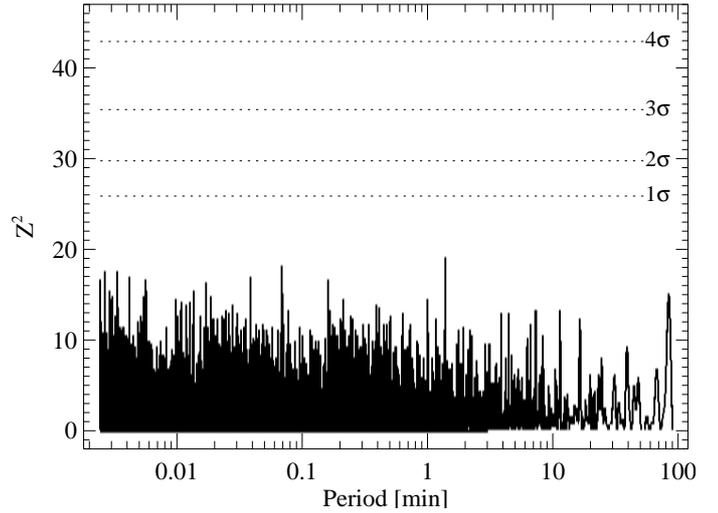}}
\caption{Results of the $Z^2$ search (EPIC-pn data, $P=0.1468-54000$\,s). 
The frequency range is $\Delta\nu\sim6.8$\,Hz, the energy band is $0.2-7$\,keV and the size of the extraction region 
is $10\arcsec$, amounting to $\sim13,500$ counts. Dashed horizontal lines show confidence levels 
of $1\sigma$ to $4\sigma$ for the detection of a periodic signal, given the frequency range, the duration 
of the observation and the number of source photons in the search. For 
plotting purposes, a logarithmic step in frequency was adopted, resulting 
in a smaller number of frequency trials than in the original search (of 
the order of $3.4\times10^6$ data points). No significant periodicity 
with pulsed fraction higher than $\sim5\%$ ($1\sigma$) is found.  
}
\label{f:z2s}
\end{figure}

\subsubsection{XMM-Newton timing analysis}

\citet{thor+09} suggested a possible magnetic nature of V405 Peg, the subclass however remained 
unclear. One of the reasons for obtaining XMM-Newton X-ray observations was the search for 
periodic X-ray variability, in particular to search for the spin periodicity of the white 
dwarf, should the object belong to the IP subclass. 

If V405 Peg would be a typical IP, the spin period could be expected in the range 
$0.05 \leq P_{\rm spin}/P_{\rm orb} \leq 0.15$ (see \citet{norton+08} for theoretical simulations 
and \citet{scaringi+10} for a recent compilation of observed data).
The orbital period of V405 Peg is 4.2635\,hr, therefore we expect $P_{\rm spin}$ to be 
between 12 min and 40 min. 
Using the Lomb-Scargle technique, we searched for a periodic behaviour of the source 
in a range between 10\,min and 40\,min. Several versions of binned EPIC-pn light curves
were used for this exercise, the original one and de-trended versions. 
De-trending was achieved by either subtracting boxcar averages or by 
subtracting the phase-averaged orbital mean. Different time bins (10\,s and 60\,s)
and different boxcar sizes were tested, 10\,m, 30\,m and 60\,m.
No significant periodicity was found in either attempt, 
and the highest peaks in the periodogram hardly reach the 
$1\sigma$ confidence limit (see Fig.~\ref{f:scargle} for the 
periodogram obtained after subtracting a 60\,m boxcar average
as one example).

Alternatively, we searched for the spin period of the white dwarf by means of 
a $Z^2_1$ (Rayleigh statistics) test \citep{buccheri+83}, which works directly 
on the (barycenter-corrected) times-of-arrival of individual 
photons. Due to high background activity, we discarded the first $\sim12$\,ks 
of the pn exposure to avoid contamination and aliasing artifacts in the search. 
In order to optimize the signal-to-noise ratio of the data, we tested several 
different energy bands and radii of the source extraction region in the period 
range of $P=0.15$\,s to 90\,min. We adopted a step in frequency of $2\mu$\,Hz 
(or an oversampling factor of 10), warranting that a peak corresponding to a 
periodic signal is not missed. Despite all tests, no significant periodicity 
with pulsed fraction higher than $5\%$ ($1\sigma$, $0.2-7$\,keV) was found 
in the period range of the search (Fig.~\ref{f:z2s}).

\begin{figure}
\resizebox{\hsize}{!}{\includegraphics[angle=-90,clip=]{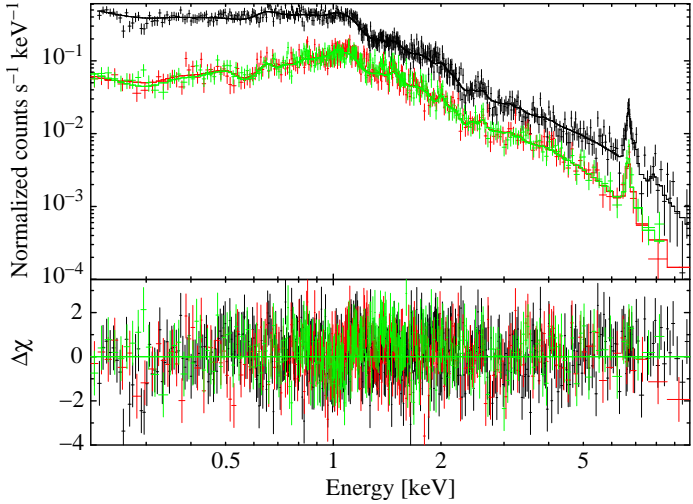}}
%[angle=270,bbllx=70pt,bblly=20pt,bburx=565pt,bbury=705pt,clip=]
\caption{Mean orbital X-ray spectrum of V405 Peg (EPIC-pn and -MOS) with 
best-fitting MKCFLOW model superposed, using MEKAL components.}
\label{f:xsp}
\end{figure}

\subsubsection{XMM-Newton X-ray spectral analysis\label{s:spec}}

The X-ray spectra were analyzed using the HEASARC software package \texttt{xspec} (version 12.7.0).
EPIC-pn and EPIC-MOS data were treated simultaneously. 
The spectra exhibit a continuum component with pronounced helium- and hydrogen-like Iron lines superposed.
The most simple model for this type of spectrum which is often encountered in CVs consists of a 
single-temperature thin thermal plasma model (MEKAL or APEC in XSPEC terminology).
Initially an absorption model was included to account for absorption by neutral interstellar matter.
The amount of hydrogen column absorption that was found was of order $N_{\rm H} = (4 \pm 3) \times 10^{19}$\,cm$^{-2}$,
hence much below the total Galactic HI column density of $N_{\rm H} = 6.15 \times 10^{20}$\,cm$^{-2}$ \citep{nh}
due to the proximity of the source. The value of this badly constrained component had no 
effect on the other parameters and we omitted the absorption component when compiling the
spectral parameters listed in Tab.~\ref{t:xfit}.
We used the elemental abundance tables by \cite{andersgrevesse89} and set those to the solar value.

The single-temperature model, however, did not provide a good fit 
to the data ($\chi^2_{\nu} = 1.6$ for 1062 degrees
of freedom), it left large positive residuals at low energies (fit parameters are listed in Tab.~\ref{t:xfit}).
A much better fit was achieved by taking a second MEKAL component into account. Formally, the fit was 
satisfactory ($\chi^2_{red}$ = 1.09 for 1060 degrees of freedom), but some coherent photon excess above 5 keV
could be indicative of a further high-temperature component. Some support for this view may be derived from 
a single-temperature fit to only the high energy range, 3-10 keV. The fit revealed $kT = 6.0 \pm 0.9$\,keV, gave a satisfactory
fitting statistics ($\chi^2_{red} = 1.05$ for 183 degrees of freedom) and left no systematic residuals at high energies.
A further improvement to the fit of the whole spectrum could thus be achieved 
when adding a third MEKAL component.
The result of this fit cannot be distinguished from the fit labeled 2MEKAL in Tab.~\ref{t:xfit}
by a comparison of the fitting statistics but the fitting residuals show random scatter only.

The sum of three MEKAL components is a rough approximation of a model with a continuous 
temperature distribution. Such a model is also available in XSPEC \citep[MKCFLOW, ][originally
designed to describe cooling flows in clusters of galaxies]{mkcflow88}, its application to V405 Peg 
yields $\chi^2_{red} = 1.06$ for 1066 degrees of freedom (fit parameters in Tab.~\ref{t:xfit}).
Figure \ref{f:xsp} shows the mean orbital X-ray spectrum together with the best-fitting MKCFLOW model.
The MKCFLOW model fits lower and upper temperature limits to an observed spectrum. We regard 
the fitted parameters as well as the associated errors as indicative only 
due to the limited spectral coverage of the EPIC detectors. 
This prevents one from reliably determining higher and lower temperatures. 
Otherwise, the measured upper limit temperature could be used to 
constrain the mass of the white dwarf.

\begin{table*}[t]
\caption{Fit statistics and fit parameters of the spectral models applied to the mean orbital 
EPIC X-ray spectrum of V405 Peg. The given errors are valid for a 90\% confidence level.}
\label{t:xfit}     
\centering                          
\begin{tabular}{lccccc}       
\hline\hline                
Model      & $\chi^2_{red}/d.o.f.$  &  $kT_{1/min}$ & $kT_2$          & kT$_{3/max}$          & F$_{2-10}$ \\   
               &                                   &  [keV]      & [keV]    & [keV]                         & [$10^{-13}$\,erg cm$^{-2}$ s$^{-1}$] \\
\hline 
MEKAL     & 1.60/1062 & $3.39 \pm 0.08$  &                         &                           & $7.2\pm 0.4$ \\    
2MEKAL   & 1.09/1060 & $0.69 \pm 0.07$  &  $3.9\pm 0.1$  &                          & $7.3\pm 0.4$ \\
3MEKAL   & 1.04/1058 & $0.65 \pm 0.03$  &  $2.1 \pm 0.3$ & $6.6 \pm 1.4$  & $7.7 \pm 0.4$ \\
MKCFLOW& 1.06/1066 & $0.24 \pm 0.08$   & ---   & $10.3 \pm 0.6$ & $7.9 \pm 0.4$ \\    
\hline                              
\end{tabular}
\end{table*}

Table~\ref{t:xfit} lists in the last column the flux in the $2-10$\,keV band, which is only 
little dependent on the chosen model. 
The bolometric correction from the unabsorbed 2 - 10 keV flux to the total flux 
for the best-fitting MKCFLOW model was determined within XSPEC by integrating 
the corresponding models and was found to be $k_{\rm bol} \simeq 2.6$.

As described in Sect.~\ref{sss:xuvlc} (see also Fig.~\ref{f:crhr}) a slight dependence of the spectral hardness
as a function of the count rate is observed.
Therefore X-ray spectra for the `bright' and the `faint' phases were extracted to search for a possible 
variability of the spectral parameters. A combined fit was applied to the two spectra simultaneously 
with a sum of two MEKAL models, a hot and a cold one.
The temperatures and the abundances of both the hot and the cold components were forced
to be the same for the bright- and faint-phase spectra, only the normalization of the components
were allowed to vary independently. Again, the fit was satisfactory, $\chi^2_{red} = 1.00$ (for 
800 d.o.f.). The temperatures of the combined fit were the same as those of the 2MEKAL fit 
applied to the mean spectrum (Tab.~\ref{t:xfit}). The normalization parameters of the combined fit are
N$_{b,h} = (8.0\pm 0.3) \times 10^{-4}$\,cm$^{-5}$,
N$_{f,h} = (7.3\pm 0.2) \times 10^{-4}$\,cm$^{-5}$,
N$_{b,c} = (4.2\pm 0.8) \times 10^{-5}$\,cm$^{-5}$,
N$_{f,c} = (3.8\pm 0.8) \times 10^{-5}$\,cm$^{-5}$,
where subscripts $b,f$ and $h,c$ denote bright and faint phases, and hot and cold 
components, respectively,  and the errors are given for a 90\% confidence interval.
Hence, the normalization of the hotter thermal component can be made responsible 
for the observed small spectral variability, a possible phase-dependent temperature change
could not be inferred from the data.

The iron line complex consists of only two components that can clearly be resolved: 
the two plasma emission lines of helium-like iron line at 6.7 keV and of hydrogen-like iron 
line at 6.97 keV. The fluorescent K$\alpha$ line of neutral iron at 6.4 keV is not detected. We estimated 
an upper limit for its line flux and the corresponding equivalent width by performing a multi-temperature 
plasma fit (MKCFLOW) to the restricted energy range of $5-8$\,keV with a superposed 
Gaussian line at fixed line energy (6.41\,keV) and line width. An unresolved line 
was assumed, the intrinsic line width was set to zero, hence the observed line width 
was defined by the instrumental resolution at this energy.
The maximum flux of the Gaussian 
line was estimated from the 90\% confidence error of the normalization parameter, 
$F_{\rm K_\alpha} = 6\times 10^{-15}$\,\fint, and resulted in an upper limit for the equivalent width
of {\ 90}\,eV.

One often finds a prominent soft component in the X-ray spectra of 
MCVs, both of the polar and the IP subtype. This is typically modeled assuming blackbody emission.
An upper limit flux for such a component was estimated by assuming cold interstellar absorption with 
$N_{\rm H}$ fixed at $5\times 10^{19}$\,cm$^{-2}$ and adding a blackbody to the MKCFLOW model with
fixed temperature $kT = 30$\,eV and was found to be $F_{\rm bol} = 1 \times 10^{-13}$\,\fint.
Hence, a soft component with a temperature as typically found in magnetic CVs reveals a bolometric 
flux which is below 5\% of the main plasma component.

\subsubsection{Swift X-ray spectroscopy}
\label{s:swspec}
Individual observations with Swift were short and suffer from low photon numbers.
Those with the highest number of photons however are not compatible with a single 
temperature plasma model. We therefore decided to use the MKCFLOW model with the 
same minimum and maximum temperature as derived from XMM-Newton and to perform 
a fit using only the normalization as free parameter. This gave a roughly satisfactory fit to 
most of the data. % ($\chi^2_{\rm red} = 0.95$ for 12 d.o.f). 
We used this model to compute a conversion factor from count-rate to bolometric flux of 
$\rm ECF = 5 \times 10^{-11}$\,ergs cm$^{-2}$ per count.
Using this ECF and assuming that the spectral shape doesn't change with brightness 
the observed variable count-rate translates into bolometric flux. All measured count-rates 
and the derived fluxes are listed in Tab.~\ref{t:xrat}.

\subsection{Optical variability} 
During the four nights of observations in July 2012, V405 Peg varied 
between $g=16.1-16.6$ on time scales as short as $\sim$$10-30$\,min 
($\sim$$0.05-0.1$ orbital phase units). According to the low- and high-state 
magnitudes listed in Sect.~\ref{s:aobs}, the July 2012 photometry clearly represents a high 
state of the object. 

The variability pattern does not show any obvious periodicity. A periodogram of these 
data derived with the analysis of variance (AOV) method \citep{schwarzenberg89} reveals
some weak signals at periods that were also present in the high-state I-band photometric 
data of September/October 2005 in \citep[][their Fig.~8]{thor+09}. In the current periodogram
one may recognize some power in the prominent period 'P2' from 2005 at 251\,min 
and its one day alias at 213\,min. The power on the orbital period is much weaker, but neither
of the mentioned periods is regarded as significant given the relatively brief observation
performed in July 2012. The high-state $g-$band variability pattern 
thus shows some resemblance to the high-state $I$-band variability.
%_______________________________________________________________________________________________________________

\begin{figure*}[t]
\resizebox{\hsize}{!}{\includegraphics[clip=]{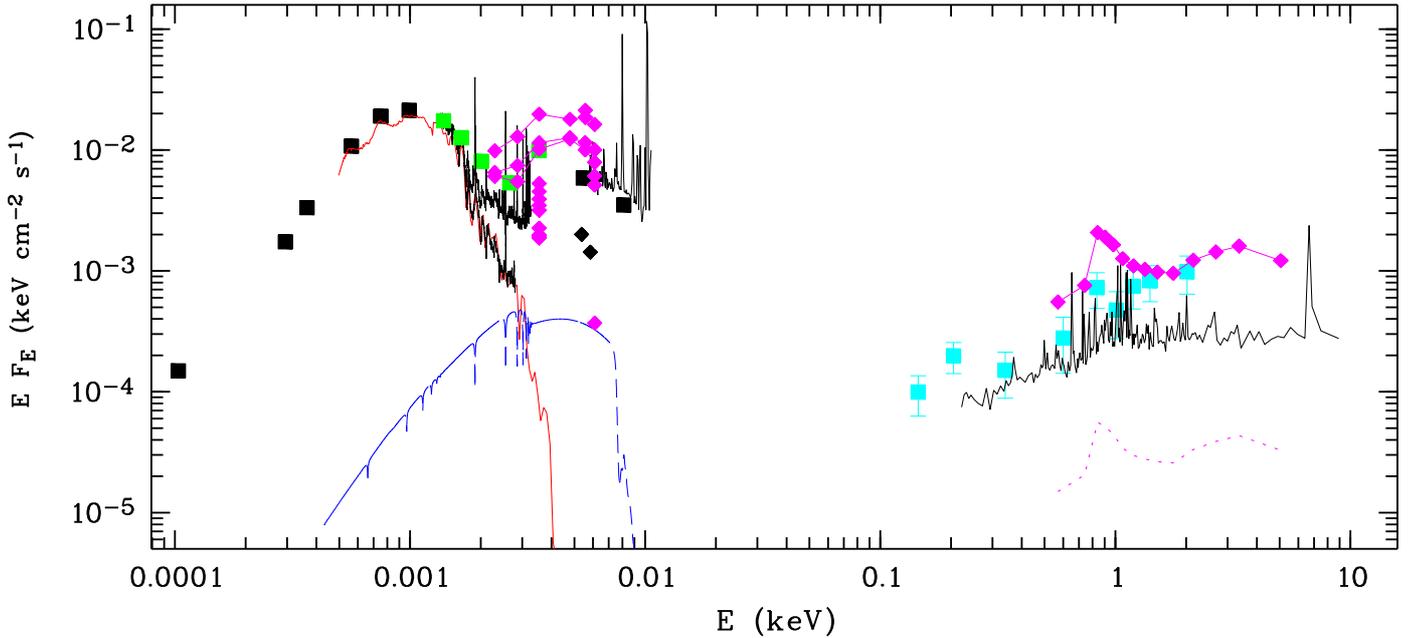}}
\caption{Infrared to X-ray spectral energy distribution of V405 Peg.
Photometric data were obtained with WISE and 2MASS (black filled squares), the SDSS 
(green squares), UVOT (magenta rhombs, data obtained through one OBSID are
connected by lines), the OM (black rhombs), and GALEX (black squares). Observed spectra 
are obtained from the SDSS (high state optical), the HST archive (high-state ultraviolet), 
and from \citet{thor+09} (low-state optical). For model spectra see text. In the X-ray spectral 
regime spectra are shown obtained with XMM-Newton (black line), Swift (magenta, high state 
spectrum obtained 2012-10-30, low state 2008-06-04; for the latter spectral 
shape was not measured, but assumed to be the same as in the high state for illustration purposes), 
and ROSAT (cyan squares). 
}
\label{f:uvirsed}
\end{figure*}

\subsection{The spectral energy distribution\label{s:sed}}

Our own follow-up and archival data led to several new data points  
to be included in the spectral energy distribution \citep[a previous version was shown by][their Fig.~11]{thor+09}. 
The updated infrared to X-ray distribution is shown 
in Fig.~\ref{f:uvirsed}. From long to short wavelengths it includes data obtained with 
WISE, 2MASS, the SDSS (both photometrically and spectroscopically), Swift-UVOT, GALEX, XMM-Newton OM,
the HST, ROSAT-PSPC, Swift-XRT, and XMM-Newton EPIC. 
Suitably adapted model spectra for the companion star and the
white dwarf are indicated with red and blue lines, respectively. They represent a cool star 
PHOENIX and a DA WD model spectrum kindly made available by Drs. Hauschildt and Koester. 

Using the recent UVOT data one can derive improved constraints for the 
white dwarf compared to \cite{thor+09}. They estimated the white dwarf's effective temperature 
from the spectral flux at the blue end of the low-state spectrum. The contribution of
the companion star however is non-negligible. We therefore firstly subtract the likely 
contribution of an M4$\pm0.5$ companion star which leaves an average 
flux of $(1.1 \pm 0.5)\times 10^{-16}$\,\flam\ ($450 - 480$\,nm). 
Assuming a standard 0.6\,\msun\ white dwarf the effective temperature 
is in the range $9500\pm1500$\,K for an assumed distance of 150\,pc. This still 
may be regarded as an upper limit flux because it ascribes the remaining flux 
to the white dwarf exclusively. It neglects a weak contribution from a recombination 
continuum, a recombination component is obvious otherwise 
through Balmer H$\alpha$ and H$\beta$ emission.
The chosen white dwarf model nicely fits to the lowest UVOT-flux obtained 
through filter UVW2 on June 4, 2008, hence the likely temperature of 10000\,K
is the maximum temperature for a standard white dwarf at the nominal distance. 

At most occasions the ultraviolet flux is a factor 10-20 higher than predicted by the 
white dwarf alone due to accretion-induced radiation. 
The UVOT-photometry traces the recombination spectrum originating 
from an accretion disk or an accretion stream from the high-state SDSS- to the 
high-state HST-spectrum. The OM, hence XMM-Newton observations, were obtained 
in an intermediate state of accretion with reduced UV flux by a factor $\sim$3 with respect to 
GALEX and most UVOT data. Our own $g$-band photometry is not shown in the figure, the 
object varied between the levels given by the SDSS-spectrum and the SDSS-photometry.

The infrared part of the SED is dominated by the companion star. The observed colors 
$W1-W2=0.163\pm0.045$ and $W2-W3=0.31\pm0.17$ are consistent with those of 
a mid M-star \citep[cf.][]{kirkp+11} leaving little (if any) room for any other radiation 
component that is prominent in other CVs (cyclotron, accretion disk, dust disk, recombination). 

The UV spectrum shows a rich emission line spectrum of HeII, CIV, SiIV, CII, Si II/III, and NV on top of a mildly red continuum.
The set of emission lines is typical for optically thin gas from either an accretion stream in 
polars or from a quiescent accretion disk in dwarf novae  \citep[for UV spectra of polars see e.g.][]{araujo+05}. 
The COS spectra of the dwarf novae DT Oct and BB Dor are almost carbon copies of that of V405 Peg. 
Hence, the average UV-spectrum does not seem to disentangle between a magnetic and non-magnetic 
nature of V405 Peg.

\begin{figure*}[t]
\resizebox{\hsize}{!}{\includegraphics[clip=]{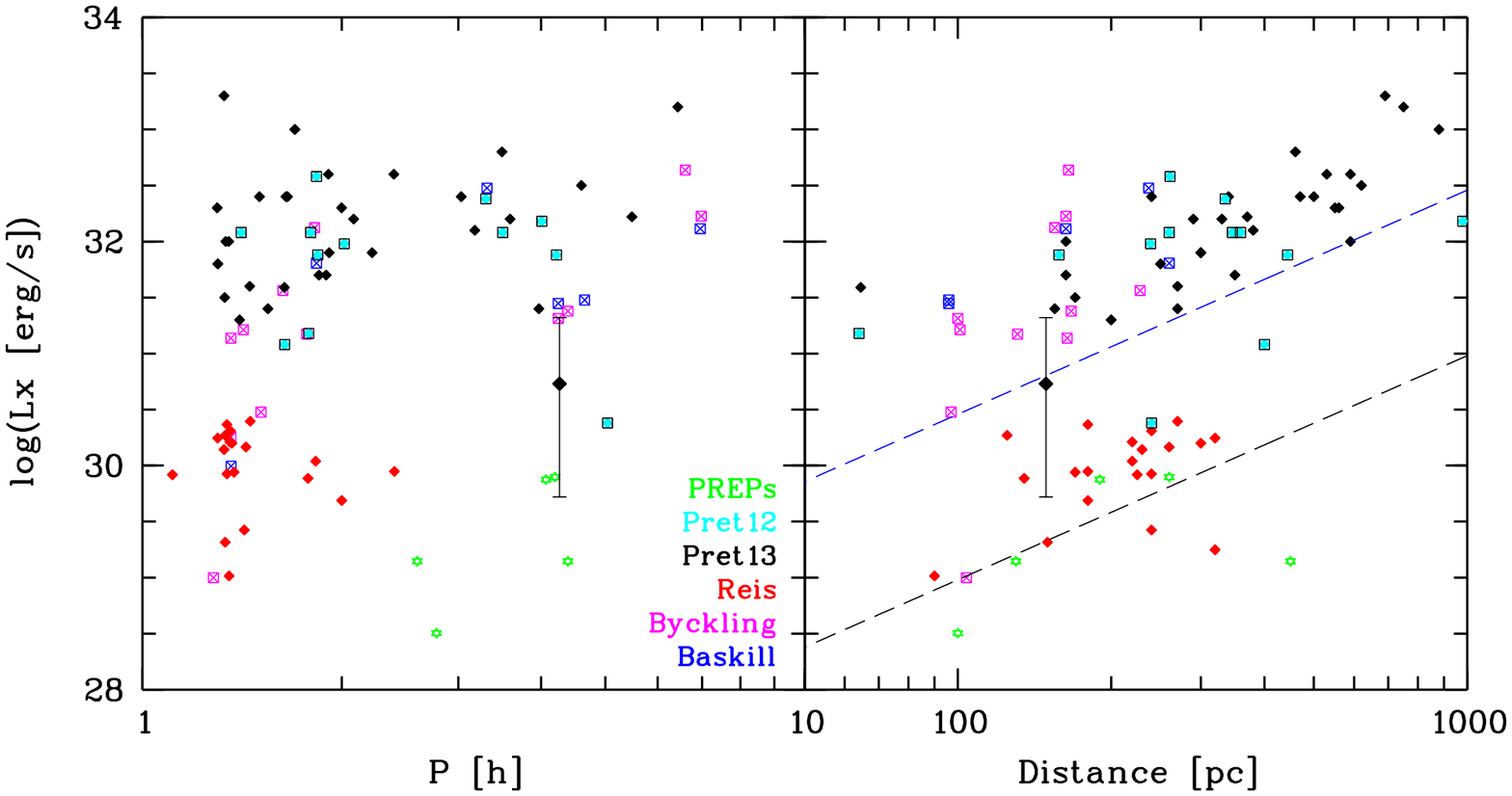}}
\caption{Magnetic and non-magnetic CVs in the 
$L_{\rm X} - P_{\rm orb}$ and the $L_{\rm X} - D$ planes. 
Luminosities, periods and distances were taken from 
\cite{pret+12}, \cite{pret+13}, \cite{reis+13}, \cite{byck+10}, and 
\cite{baskill+05}. Luminosities given in certain spectral bands were corrected 
with bolometric correction factors using the spectral parameters 
given in the papers mentioned.
The black larger symbol indicates the XMM-Newton observation of V405 Peg, whereas the bars
crossing it indicate the range of luminosities as observed with Swift. Diagonal 
lines in the right panel indicate the survey limits of the RBS and with 
eROSITA.
}
\label{f:lxporb}
\end{figure*}

\section{Discussion}

We have analyzed new original X-ray and optical spectral and timing 
data of the peculiar low-luminosity cataclysmic variable V405 Peg and 
put them in context with multi-lambda data available in the literature. 

{\it What kind of CV is V405 Peg?} The various possibilities were discussed by \cite{thor+09} and 
we re-iterate here the main points in the light of the new observations. Evidence may 
be found from the X-ray spectrum, X-ray and optical variability on orbital and much longer
time scales, and the spectral energy distribution.  
 
The X-ray spectrum is well reflected by an optically thin multi-temperature plasma which accounts 
for both the continuum and the ionized emission lines of Fe. That kind of X-ray spectrum
is found in magnetic and non-magnetic CVs, originating either from the post-shock regions 
in accretion columns or the boundary layers of disk-accreting CVs. The absence of a black-body like
soft X-ray component does not further help to classify the object, since the notorious soft X-ray component 
of many ROSAT-discovered magnetic CVs can no longer be regarded as their observational hallmark \citep{ramsaycropper04}.
The XMM-Newton X-ray spectrum does not show any sign of absorption, neither by cold interstellar matter
nor by warm matter in the vicinity of the site where the X-rays are generated. There is also no sign 
of X-ray reflection, but again, this does not classify the source, since Fe reflection lines were seen
in magnetic and non-magnetic CVs. 

Perhaps most constraining at X-ray wavelengths is the observed variability, both on short (orbital) and 
longer timescales. V405 Peg is a low-inclination system, orbital X-ray variability in non-magnetic CVs
is observed only in high-inclination system \citep{baskill+05}, hence the observed orbital variability 
argues for a magnetic cataclysmic variable. Also the observed large-amplitude variability at X-ray 
and ultraviolet wavelengths gives further evidence for the absence of an accretion disk which 
would act as a mass repository. The X-ray luminosity varied between $5 \times 10^{29}$\,\lum\ 
and $2 \times 10^{31}$\,\lum\ among the Swift pointings. When observed with XMM-Newton the 
object was in an intermediate state of accretion with $5 \times 10^{30}$\,\lum, which compares 
to its brightness at discovery in the RASS. If this brightness is typical, the object is under-luminous at the 
given orbital period, at its bright limit it roughly compares to other CVs at that period (see 
Fig.~\ref{f:lxporb}  and the discussion further down in this section).

If magnetic, a further subdivision into the polar or the intermediate polar subclass would be of interest.
Neither our new time-resolved optical photometric observations nor the X-ray data 
revealed any significant (new) period. We noticed weak X-ray variability as a function 
of the binary orbital phase. The absence (or better saying the non-detection) 
of a much shorter periodicity that could be associated with the spin of the white dwarf, might 
argue for a CV of polar type, but evidence is low.
The X-ray bright phase is centered roughly on phase zero, which would fit in the 
known picture of polars where the accreting pole faces the mass-donating star.  
If being a polar, the non-detection of any magnetic feature in the optical/UV/IR is puzzling.
Neither Zeeman-shifted white-dwarf atmospheric absorption lines nor cyclotron emission lines
could be discovered. The observed SED leaves little room for a pronounced cyclotron component 
at all. However, a final statement on the presence of a weak cyclotron component requires 
time-resolved multi-wavelength polarimetry, which is encouraged.

We could further constrain the temperature of the white dwarf in the system through 
UVOT and OM photometry to be likely below 10000\,K (the previous limit was 17000\,K), 
which makes membership of the VY Sculptoris subclass even more unlikely than 
discussed previously. The UV emission line spectrum obtained in a high state 
shows similarities with both quiescent dwarf novae and normally accreting polars. 
The observed very low temperature of the WD might give further evidence for a magnetic CV. 
Similarly cool WDs were found in long-period CVs, i.e.~those above the $2-3$\,hour period gap,  
in polars only.  There is not any other WD
in a long-period CV among the objects studied by \cite{townsleygaensicke09} that is as 
cool as that in V405 Peg (see in particular their Fig.~5). 
The implied very low mean mass transfer rate, $\langle\dot{M}\rangle < 10^{-10}$\,\msun yr$^{-1}$ 
is orders of magnitude below that expected for traditional magnetic braking \citep{howell+01}  
but merely consistent with gravitational radiation losses only. 
Our result rests on the low-state spectrum (spectrophotometric uncertainty $\sim$20\%)
and the one low-state photometric data point obtained with Swift-UVOT.
Hence, a low-state, phase-resolved, spectroscopic study in the ultraviolet 
would therefore be very much demanding, to ascertain our tentative result and thus 
confirm the existence of a long-period CV with a mean mass transfer rate below all 
predictions of wind-driven angular momentum loss models.

{\it What are the prospects of finding more objects of the V405 Peg type?}
It might be instructive to place V405 Peg in the $L_{\rm X} - P_{\rm }$ and the
$L_{\rm X} - D$ landscapes of cataclysmic 
variables. In Fig.~\ref{f:lxporb} various CV subsamples are shown with well-determined 
distances, part of those were used to determine the space density and luminosity 
functions of CVs. Data were taken from \citet{baskill+05,reis+13,pret+12,pret+13,byck+10}. 
\cite{byck+10} were using a sample of non-magnetic CVs (dwarf novae, to be more precise)
with parallax-based distance measurements to constrain the CV luminosity function. 
\cite{baskill+05} presented the complete set of ASCA observations of non-magnetic CVs, 
only a subset has precise distances, hence luminosities. \cite{reis+13} studied a sample of 
optically selected (mostly SDSS) non-magnetic CVs which they followed with Swift to uncover 
the intrinsic differences in the X-ray properties of optically selected CVs and those that were discovered
via other routes. \cite{pret+12} and \cite{pret+13} were using X-ray selected CVs (mostly from the 
RBS) to derive space densities and uncover intrinsic differences between magnetic and non-magnetic 
CVs. 
Data for the (likely) wind-accreting pre-polars \citep[PREPs, ][]{schwope+09} -- 
often referred to a LARPs \citep[Low-Accretion Rate Polars][]{schwope+02}  --  
are also included in the figure. They were adopted 
from \cite{vogel+07,vogel+11} and yet unpublished own results. 

\cite{reis+13}, \cite{pret+12} and \cite{pret+13} 
each used a certain spectral model with fixed parameters to convert the 
measured count rates with Swift and ROSAT to fluxes and luminosities.  
These were derived for the spectral ranges covered by the mentioned satellite observatories. 
The X-ray luminosities as given there were thus  
modified by bolometric correction factors of $\Delta L=0.17$ and $0.68$, respectively, to 
put all luminosity values that are displayed in the diagram on a common scale. The bolometric correction factors
were computed using XSPEC with the spectral parameters 
as given in the mentioned papers.

Double entries were removed from the 
various lists of CVs. The diagonal lines in the right panel 
indicate the flux limits of the ROSAT Bright Survey \citep[RBS,][]{schwope+00}, 
$f_X= 2.4 \times 10^{-12}$\,erg\,cm$^{-2}$\,s$^{-1}$,
and that of a corresponding eROSITA bright survey, which will
be a factor 30 more sensitive \citep{merloni+12}. 

Accreting white-dwarf main-sequence binaries are seen to occupy distinct regions in the 
diagrams. Outbursting and intrinsically bright objects are observed at all periods 
with $\log$\lxbol\, $>30.5$\,\lum. 
Those objects show a correlation between 
orbital period and X-ray luminosity \citep{baskill+05}. This appears, however, 
much weaker than expected from population synthesis, would the X-ray luminosity 
scale proportionally with the mass accretion rate \citep{howell+01}.

In contrast to this (seemed) typical CV population,
\cite{reis+13} recently presented an X-ray study of SDSS-selected, 
short-period, white-dwarf dominated, hence low-luminosity objects. 
About 90\% of the objects studied 
by them were found to be X-ray emitters, all of them were found below the RBS-limit. 
In particular, they suggest that such sources could contribute significantly 
to the GRXE (Galactic Ridge X-ray Emission), would this (yet to be discovered)
large population of low-luminosity objects exist.
An optical selection based on colors for spectroscopic follow-up in 
the SDSS removes an X-ray luminosity bias from existing samples, however, 
it introduces a color-dependent bias. Since all CVs are known to be notorious 
X-ray emitters, unbiased samples may eventually be constructed from sensitive X-ray surveys,
like those planned with eROSITA. 

Even fainter but at longer orbital periods than the SDSS-selected CVs studied by \cite{reis+13} 
are the wind-accreting PREPs. 
Contrary to the other objects shown in the diagram they are thought 
to be Roche-lobe under-filling, which explains their low luminosities. 
V405 Peg at a long orbital period, i.e.~above the period gap, 
with its huge observed X-ray variability represents a bridge 
between the PREPs and the CVs accreting from Roche-lobe overflow.

V405 Peg may be the first of many more to be discovered from eROSITA surveys. 
With ROSAT, similar objects could be detected within a radius of 150\,pc. 
If as bright as during the XMM-Newton observations analyzed in this paper,
such objects will be detectable up to 800\,pc, sufficient to compose 
volume-limited samples which shall be essentially free from selection bias.

V405 Peg is just one out of 42 CVs studied by Pretorius et al.~(2012, 2013) above the
flux limit of the RBS and thus might be regarded as rare. It is worth to be noted 
that on the other hand one (EX Dra) out of 4 CVs discovered in the much deeper NEP survey
(North Ecliptic Pole survey) shares properties 
with V405 Peg ($P_{\rm orb} = 5.039$\,h, $\log$\lxbol$ = 30.4$\,\lum; EX Dra is the 
object next to V405 Peg in Fig.~\ref{f:lxporb} at slightly longer period). The upcoming 
eROSITA surveys will be similarly deep as the NEP survey but cover the whole sky, not just 
81 square degrees. 

\section{Conclusions}
We have analyzed new observations obtained with XMM-Newton and Swift
at X-ray and ultraviolet wavelengths, and optical photometry with STELLA/WiFSIP
and discussed our results together with multi-wavelength archival observations.
The nature of V405 Peg remains elusive in the sense that its CV subclass remains uncertain. 
It may represent the first of a large population of yet to be discovered almost 
Roche-lobe filling CVs by future sensitive X-ray surveys.

\begin{acknowledgements}
We thank our referee for constructive criticism which helped to improve the paper.

This work has been supported by the Deutsches Zentrum f\"ur Luft-
und Raumfahrt under contracts 50 0R 1011 and 50 OX 1101.

AP acknowledges support by the Deutsche Forschungsgemeinschaft under 
grant Pi983/1-1.

We thank M.J.~Page for providing the UVOT photometric data.

This research has made use of the NASA/ IPAC Infrared Science Archive, 
which is operated by the Jet Propulsion Laboratory, California Institute of 
Technology, under contract with the National Aeronautics and Space Administration.

Some of the data presented in this paper were obtained from the Mikulski Archive for 
Space Telescopes (MAST). STScI is operated by the Association of Universities for Research in 
Astronomy, Inc., under NASA contract NAS5-26555. Support for MAST for non-HST data is 
provided by the NASA Office of Space Science via grant NNX09AF08G and by other grants and contracts.

Funding for the SDSS and SDSS-II has been provided by the Alfred P. Sloan Foundation, 
the Participating Institutions, the National Science Foundation, the U.S. Department of Energy, 
the National Aeronautics and Space Administration, the Japanese Monbukagakusho, the Max 
Planck Society, and the Higher Education Funding Council for England. The SDSS Web Site is 
http://www.sdss.org/.

The SDSS is managed by the Astrophysical Research Consortium for the Participating Institutions. 
The Participating Institutions are the American Museum of Natural History, Astrophysical Institute 
Potsdam, University of Basel, University of Cambridge, Case Western Reserve University, University 
of Chicago, Drexel University, Fermilab, the Institute for Advanced Study, the Japan Participation 
Group, Johns Hopkins University, the Joint Institute for Nuclear Astrophysics, the Kavli Institute 
for Particle Astrophysics and Cosmology, the Korean Scientist Group, the Chinese Academy of 
Sciences (LAMOST), Los Alamos National Laboratory, the Max-Planck-Institute for Astronomy (MPIA), 
the Max-Planck-Institute for Astrophysics (MPA), New Mexico State University, Ohio State University, 
University of Pittsburgh, University of Portsmouth, Princeton University, the United States Naval 
Observatory, and the University of Washington.
\end{acknowledgements}

\bibliographystyle{aa}
\bibliography{aa22662}
  
\end{document}